\documentstyle[emulateapj]{article}

\newcommand{\be}{\begin{equation}}
\newcommand{\ee}{\end{equation}}
\newcommand{\ba}{\begin{array}}
\newcommand{\ea}{\end{array}}
\newcommand{\siml}{\lower4pt \hbox{$\buildrel < \over \sim$}}
\newcommand{\simg}{\lower4pt \hbox{$\buildrel > \over \sim$}}
\def\Mesz{M\'esz\'aros~}

\slugcomment{Accepted for publication in ApJ}

\begin{document}

\title{Gamma-ray burst beaming: a universal configuration with a
standard energy reservoir?}

\author{Bing Zhang \& Peter M\'esz\'aros}
\affil{
Department of Astronomy \& Astrophysics, Pennsylvania State
University, University Park, PA 16803
}

\begin{abstract}
We consider a gamma-ray burst (GRB) model based on an anisotropic
fireball with an axisymmetric energy distribution of the form
$\epsilon(\theta) \propto \theta^{-k}$, and allow for the observer's
viewing direction being at an arbitrary angle $\theta_v$ with respect
to the jet axis. This model can reproduce the key features expected
from the conventional on-axis uniform jet models, with the novelty
that the achromatic break time in the broadband afterglow lightcurves
corresponds to the epoch when the relativistic beaming angle is equal
to the viewing angle, $\theta_v$, rather than to the jet half opening
angle, $\theta_j$.  If all the GRB fireballs have such a similar energy
distribution form with $1.5 < k \siml 2$, GRBs may be modeled by a
quasi-universal beaming configuration, and an approximately standard
energy reservoir. The conclusion also holds for some other forms of
angular energy distributions, such as the Gaussian function.
\end{abstract}

\keywords{gamma rays: bursts - shock waves - ISM: jets and outflows}

\section{Introduction}
Recently, several independent approaches have led to the conclusion
that long gamma-ray bursts (GRBs) have a standard energy reservoir
of several $10^{50}~{\rm ergs}$ (Frail et al. 2001, hereafter F01;
Panaitescu \& Kumar 2001, hereafter PK01; Piran et al. 2001, hereafter
P01). An important ingredient of this argument is that the putative
jet opening angles, $\theta_j$, as inferred from the afterglow lightcurve
breaking times, $t_b$, have a broad distribution, but just of the right
form to compensate for the wide dispersion of the ``isotropic'' energy
emitted in $\gamma$-rays, $E_{\gamma,{\rm iso}}$, so that $E_\gamma=
(E_{\gamma,{\rm iso}}/4\pi) (\theta_j^2/2)$ is essentially invariant
(F01). The total energy of the fireball should be $E_{\rm tot}
\geq E_\gamma +E_0$, where $E_0$ is the initial kinetic energy of the
fireball in the afterglow phase assuming an adiabatic evolution, and
the inequality takes into account the possible energy loss during the
radiative regime in the early afterglow phase that has evaded the
present observations, as well as energy losses outside the
$\gamma$-ray band (e.g. the BATSE window) or in non-electromagnetic
forms (e.g. neutrinos and gravitational waves) during the prompt phase.
Writing $E_\gamma=\eta E_{\rm tot}$ where $\eta$
is the gamma-ray emission efficiency, $E_{\rm tot}$ could be mainly
contributed by $E_0$ if $\eta$ is small (e.g. $<0.1$). PK01 and P01
found that $E_0$ is also distributed in a narrow range. For a uniform
jet, this leads to the inference that $E_0=(dE/d\Omega) (\theta_j^2/2)$
is also essentially invariant. However, in the above analysis, and in
the current afterglow jet models which are used to determine $\theta_j$,
it is generally assumed that the jets are uniform, with sharp cut-offs
at the edges, and that the line-of-sight cuts right across the jet axis.
None of these assumptions are necessarily true in general (\Mesz,
Rees \& Wijers 1998; MacFadyen \&  Woosley 1999; Woods \& Loeb 1999;
Nakamura 1999; Paczy\'nski 2001; Salmonson 2001; Dai \& Gou 2001). On
the other hand, although it is not difficult to construct a central
engine model which makes GRBs with a standard energy reservoir but
with quite different beaming angles, it would be more elegant to have
a model that all the GRB beams share a standard energy reservoir as
well as {\em a quasi-universal beaming configuration} (M. J. Rees,
2001, private communication). Here we show that such a model can be
constructed by taking account of the off-axis anisotropic jet effects,
without violating the present observational constraints.

\section{The Model}

Our assumption is that all the long GRBs have a quasi-universal
beam configuration, with a strong anisotropy of the angular
distribution of the fireball energy around an axial symmetry.
The jet axis is physically related to the rotational axis of
the central engine, so it is reasonable to assume that initially the
closer to the jet axis, the higher the energy concentration. The
actual angular distribution of the fireball energy is unknown, and we
model it as (e.g. \Mesz et al. 1998)
\be
dE/d\Omega=\epsilon(\theta,\phi)=\epsilon(\theta)=\epsilon_0
\theta^{-k},
\label{eps}
\ee
within the range $\theta_m\leq \theta \leq \Theta$, where $\theta_m$
is a very small angle within which some deviation from (\ref{eps})
is necessary to avoid the divergence at $\theta=0$, and $\Theta$ is
some large angle which exceeds the presently measured $\theta_j$ by at
least a factor of two (for the simplification of the discussions
below). The real angular energy distribution may differ from the
power law (\ref{eps}), but most of our discussions below can be
generalized to other forms of distribution functions (e.g. see
[\ref{Gauss}] and relevant discussions below). The adoption of
(\ref{eps}) is for the simplicity of the discussions.
The angular-dependence of the baryon loading rate is uncertain, and we
assume that it is weak
so that the Lorentz factor angular distribution follows a similar law,
i.e., $\Gamma(\theta) \propto \theta^{-k}$ (of course, the law should
be modified when $\Gamma(\theta)$ approaches unity). We make furthermore
the assumption that
\be
(1.5) < k \siml 2
\label{k}
\ee
in (\ref{eps}). The reason for this requirement will become evident
later. The main conjecture of the model is that {\em the dispersion in
the afterglow data of the breaking time, $t_b$, is a manifestation of
the diversity of viewing angles of the observers, rather than to the
diversity of intrinsic opening angles of the jets themselves.} In other
words, that what were inferred by Frail et al. (2001) as $\theta_j$ are
essentially $\theta_v$ in our model, where $\theta_v$ is the observer's
viewing angle with respect to the jet axis. We will test whether the
above hypothesis is able to pass the following three criteria: (i) When
$\Gamma(\theta_v) \gg 1/\theta_v$, the jet dynamics along the
line-of-sight satisfies the isotropic law $\bar\Gamma(\theta_v,t)
\propto t^{-3/8}$ (for simplicity, we only discuss an adiabatic
fireball running into an interstellar medium with a constant density),
where $t$ is the observer time, and $\bar\Gamma(\theta_v,t)$ is an
effective Lorentz factor assuming an isotropic fireball which could
mimic the emission in the direction $\theta_v$ at the time $t$; (ii)
When $\Gamma(\theta_v) \siml 1/\theta_v$, the dynamics changes so that
the lightcurves steepen;
(iii) The total jet energy $E_{\rm tot}$ is essentially a universal
value.

For an isotropic adiabatic fireball running into a uniform medium,
$\Gamma(t)\propto \epsilon^{1/8} n^{-1/8} (at)^{-3/8}$, where
$\epsilon=dE/d\Omega$ is the energy per solid angle, $n$ is the
ambient medium number density, and the blastwave radius is written in
a general form as $R=a\Gamma^2 ct$, where the factor $a$ effectively
takes into account the surface of equal-arrival-time as well as the
thickness of the emitting region. To test the criteria (i), the key is
to estimate the effective energy per solid angle, $\bar\epsilon
(\theta_v,\phi_v,t)=\bar\epsilon(\theta_v,t)$, in the direction
$(\theta_v,\phi_v)$, and to evaluate the possible time-dependence of
this value.
When $\Gamma(\theta_v,t)=\Gamma \gg 1/\theta_v$, the observer
can only observe a solid angle around $(\theta_v,\phi_v)$ with a half
opening angle of order $1/\Gamma$ due to the relativistic beaming
effect\footnote{Strictly speaking, the observer will see a smaller
half cone on the close side of the jet axis, and a larger half cone
on the far side to the axis, due to different Lorentz factors in
different directions. This will modify the integral limits in
(\ref{epsbar}), but does not influence the conclusion in
(\ref{epsbar2}) and the relevant discussions.}. By definition, the
effective energy per solid angle in the direction $(\theta_v,\phi_v)$
is
\be
\bar\epsilon (\theta_v,\phi_v,t)
=\bar\epsilon (\theta_v,t)
=\frac{\int_{\theta_v-1/\Gamma}^{\theta_v+1/\Gamma}
\epsilon(\theta,t)\sin\theta d\theta}
{\int_{\theta_v-1/\Gamma}^{\theta_v+1/\Gamma}
\sin\theta d\theta}
\label{epsbar}
\ee
due to the axial symmetry.
In the small angle approximation, which is relevant to the present
discussions\footnote{The largest ``jet'' angle in F01
is 0.411, and the approximation is good within 3\%.},
one has $\sin\theta\sim \theta$. When $\Gamma \gg 1/\theta_v$ and
noticing (\ref{eps}), this gives
\be
\bar\epsilon(\theta_v,t)\simeq \frac{\int_{\theta_v-1/\Gamma}^{\theta_v
+1/\Gamma}\epsilon_0 \theta^{1-k} d\theta}{\int_{\theta_v-1/\Gamma}^{\theta_v
+1/\Gamma} \theta d\theta} \simeq \epsilon_0\theta_v^{-k}=\epsilon(\theta_v).
\label{epsbar2}
\ee
This is a time-independent quantity, since the sideways expansion
effect is not important at the same stage (see discussions below). We
then have
\be
\bar\Gamma(\theta_v,t) \propto [\bar\epsilon(\theta_v,t)]^{1/8}
n^{-1/8} (at)^{-3/8} \propto t^{-3/8}, ~~\Gamma \gg1/\theta_v.
\ee
This indicates that the observer does not feel the anisotropy of the
fireball when the relativistic beaming angle $1/\Gamma$ is much
smaller than the viewing angle $\theta_v$, but observes the fireball
as if it were isotropic. This is the same conclusion as drawn in the
on-axis uniform jet model. The conclusion (\ref{epsbar2}) does not
require (\ref{k}) and holds for any $k$ value. In fact, it even holds
for some other forms of angular energy distributions, e.g. the Gaussian
distribution,
\be
\epsilon(\theta) = \epsilon_0 \exp[-(1/2)(\theta/\theta_0)^2],
\label{Gauss}
\ee
as long as the first order Taylor expansion term of these functions
lead to the same result.
Notice that the factor $a$ of various forms may deviate from the conventional
value (e.g. $\sim 4$), since the shape of the equal-arrival-time surface
will be distorted due to the anisotropic distribution of the fireball
energy. However, its time-dependence, if any, would be very
small. Therefore, although it may influence the absolute values of the
afterglow flux levels, such an effect does not change the blastwave
dynamics in the viewing direction.

In principle, the jet configuration is time-dependent, due to the
effects such as the energy redistribution and the lateral expansion.
In the lab frame, the causally connected region subtends an angle of
$\Delta\theta \sim c_s t'/R=c_s/c\Gamma(\theta)$, where
$t'=R/c\Gamma(\theta)$ is the comoving time since the explosion
along the $\theta$ direction, $c_s$ is the expansion speed, which may
be either the relativistic speed of sound $\sim c/\sqrt{3}$ (Rhoads
1999), or simply the speed of light (Sari et al. 1999).
As $\Gamma(\theta_v) \gg 1/\theta_v$, the line-of-sight direction is
causally disconnected from other regions, so the dynamics evolves
essentially independently, so that the description in (\ref{epsbar})
holds. When the blastwave decelerates so that the line-of-sight bulk
Lorentz factor $\Gamma(\theta_v)$ drops close to and below
$1/\theta_v$, the dynamics along the line-of-sight starts to change,
and $\bar\Gamma(\theta_v)$ will deviate from the $\propto t^{-3/8}$
dependence. There are several effects that play a role. First, as
$1/\Gamma(\theta_v)$ exceeds $\theta_v$, the observer starts
to feel the energy deficit due to the drop of the energy distribution,
i.e. deviation of the power law (\ref{eps}), on the other side of
the jet axis. Although the calculation of $\bar\epsilon(\theta_v,t)$
is no longer straightforward, this deficit effect should mimic that in
the uniform jet model as long as $k$ is not too flat, say, $k>1.5$.
Second, the anisotropic jet has a trend to resume the isotropic shape.
As $\Gamma(\theta_v) \sim 1/\theta_v$, the viewing direction starts to
connect the jet axis causally.  The energy outflow from the cone
defined by $\theta_v$ becomes prominent, and this equivalently
decreases $\bar\epsilon (\theta_v)$ in the viewing direction.
In the meantime, the initial material within the $\theta_v$ cone
starts to spread into a wider cone, and the observer would feel a
stronger deceleration, although the global sideways expansion will
become evident only when $\Gamma(\theta_v)$ drops below
$1/\Theta$. All these effects tend to steepen the afterglow lightcurve,
although the degree of steepening is unclear without detailed
numerical dynamical calculations. In any case, in the asymptotic
phase of sideway expansions, $R$ is essentially a constant,
and one would eventually have  $\bar\Gamma(\theta_v,t)\propto t^{-1/2}$
(from $t \sim R/\bar\Gamma^2$, Rhoads 1997, 1999; Sari et al. 1999).
In this regime, the temporal indices of the lightcurves in the various
spectral regimes would follow closely the same predictions
as in the uniform jet models (Sari et al. 1999; Rhoads 1999). For
example, in the slow-cooling regime (which is usually the case after
the viewing or ``jet'' break), for spectral regimes both below and
above the cooling frequency, the asymptotic spectral flux is $F_\nu
\propto t^{-p}$, where $p$ is the power-law index of the electron number
distribution. For reasonable values of $p$ (e.g. $\sim 2.2$), this is
consistent with several GRB afterglow observations. The above
discussion should also hold for other distribution functions such as
(\ref{Gauss}), mainly because eventually all the initial
configurations will be smeared out. However, to address the
lightcurves properly within different models, including the
relevant gradual transition between asymptotic regimes, a detailed
dynamical description and numerical calculation is necessary, and we
postpone this to a future work.

We have shown that the present model can reproduce the key features of
the on-axis uniform jet model, with an arbitrary $k$ value as long as
it is not too flat. The next question is whether the model can
also retain the merit of a standard energy reservoir invoked in the
conventional jet model. In principle, one does not have to fulfill
this constraint, but just wishes so for the sake of elegance.
By definition, the total energy in a fireball with an energy
distribution given by (\ref{eps}) is
\be
E_{\rm tot}=2\pi \int_0^{\Theta} \epsilon(\theta) \sin\theta d\theta
\simeq 2\pi \int_{\theta_m}^\Theta \epsilon_0 \theta^{1-k}d\theta.
\label{Etot}
\ee
For $k<2$ and $\Theta \gg \theta_m$, we get
\be
E_{\rm tot}\simeq \frac{2\pi}{2-k} b^{2-k} \epsilon(\theta_v) \theta_v^2,
\label{b}
\ee
where we have parameterized $\Theta=b\theta_v$. We can see that the
quantity $2\pi\epsilon(\theta_v) \theta_v^2$ (which is essentially
the $E_\gamma$ of F01, or $E_0$ of PK01 and P01) is quasi-invariant, if
$k$ and $E_{\rm tot}$ are constant (or have a small scatter). The only
extra scatter is introduced through the scatter of $b$, which is
introduced by the scatter of $\theta_v$ (assuming the same $\Theta$
for all GRBs). However, for the index $(2-k)$ this scatter is greatly
reduced if $k$ is not much smaller than 2. This is another reason why
we require, say, $k>1.5$, in (\ref{k}). A smaller $\Theta$ can also
reduce the $b$ scatter. Notice that the $b$ scatter tends to raise
$E_{\rm tot}$ in GRBs with smaller $\theta_v$'s (and hence larger
$b$'s), which seems to be helpful to reduce the $E_0$ scatter in PK01.
An important implication of equation (\ref{b}) satisfying such a
constraint is that the total energy reservoir is standard,
but the absolute value need no longer necessarily be several times
$10^{50}$ ergs, but would depend on the value of $k$ and the
typical value of $b$. Given reasonable values, $E_{\rm tot}$ could be
one order of magnitude higher than that of F01 and PK01, but this
could be still well accommodated within conventional central engine
models (\Mesz, Rees \& Wijers 1999). The closer $k$ approaches 2, the
larger the standard energy reservoir one requires. At $k=2$, equation
(\ref{b}) should be modified in a form containing a logarithmic term,
and the energy requirement is the highest (see discussions in Rossi,
Lazzati \& Rees 2001). Also the scatter of $\theta_m$ must be very
small for $k=2$, while for $k<2$, the actual value of $\theta_m$ is
not important.
For $k \geq 2$, generally $E_{\rm tot}$ (eq.[\ref{Etot}]) can not be
expressed in terms of $\epsilon(\theta_v) \theta_v^2$, since most
of the energy is distributed at small angles. The standard energy budget
argument no longer holds\footnote{However, if $\theta_m$ is not too small
(e.g. a not very small fraction of $\theta_v$), the case $k \simg 2$
could still retain the feature of a standard, finite (but even larger)
energy reservoir.}.
A quasi-universal beaming configuration as well as a standard energy
reservoir is however in general obtained if the requirement (\ref{k})
is satisfied for an energy distribution such as (\ref{eps}).

For other forms of energy distributions, a standard energy reservoir
is also attainable. For example, for the Gaussian distribution
(\ref{Gauss}), one has $E_{\rm tot} \sim \epsilon_0 \theta_0^2$.

\section{Discussion}

We have shown that an off-axis anisotropic jet with an energy
distribution with angle given by equation (\ref{eps})
(or other forms such as [\ref{Gauss}]),
is able to reproduce the key observational features of a conventional
on-axis uniform jet model, e.g. such as producing a ``jet break"
signature in the light curve.
The novelty here is that the achromatic break time $t_b$ in the
broadband afterglow lightcurves no longer corresponds to the time when
the relativistic beaming angle is equal to the jet half opening angle,
$\theta_j$. Rather, it corresponds to the time when the relativistic
beaming angle is roughly equal to the observer's viewing angle
$\theta_v$ relative to the jet axis. In this model, the broad
distribution of $t_b$ in the data is no longer due to the intrinsic
scatter of the jet opening angles among different bursts, but is
attributed to the distribution of the observer's lines of sight.
For a power law energy distribution (\ref{eps})
(or a Gaussian energy distribution [\ref{Gauss}]),
if the constraint (\ref{k}) is satisfied, all the GRBs may have a
quasi-universal beaming configuration, besides a quasi-standard energy
reservoir. We deem this to be a more elegant picture than the conventional
on-axis uniform jet model. In addition, the homogeneous nature of the
conventional model is more idealized, and the present inhomogeneous model
is likely to be a closer representation of what could be expected in nature.

The predictions of this inhomogeneous model for the afterglow lightcurves
are not completely equivalent to those of the uniform jet model. The key
difference should occur around the ``jet break'' time. Our model should
give a more gradual variation at the break than the uniform jet model,
which assumes a sharp drop off at the jet edge. The so far sparsely studied
sideways expansion effect in an anisotropic jet may further complicate
the problem. The shape of the break should also depend on the angular
energy distribution function and some unknown parameters, such as $k$.
Detailed modeling is necessary in order to address these questions.
In any case, the gradual break expected in our model is not
inconsistent with several well studied afterglow lightcurves, and some
simulations have shown that the conventional jet models usually
also give gradual and smooth jet breaks (e.g. Panaitescu \& \Mesz, 1999;
Moderski, Sikora \& Bulik 2000; Huang et al. 2000). Both models are
compatible with the present data, but this situation may change as better
data becomes available and as more detailed simulations are performed.

Recently, Rossi et al. (2001) have independently discussed the power-law
model (\ref{eps}) in more detail.
They plotted the afterglow lightcurves for the $k=2$ case which
mimic those of the on-axis uniform jet model, and also discussed the
more general cases of $k\neq 2$.
Here we have presented a general analytical argument, showing that
the blastwave dynamics at the line of sight is identical to the
uniform jet model in the asymptotic regime for a locus of models of
the general form of equation (\ref{eps}), as long as $k$ is not too
flat. With this particular form of the angular dependence of the energy,
in order to have a standard, finite energy reservoir for all bursts one
requires the constraint (\ref{k}). The upper end $k \siml 2$ of the
constraint (\ref{k}) ensures that the total energy can be expressed
in terms of $\epsilon(\theta_v) \theta_v^2$ and does not diverge.
(However, the case $k \simg 2$ could also have the same virtue if
$\theta_m$ is not too small compared with $\theta_v$).
The lower end of the constraint, $k > 1.5$, ensures that the scatter
introduced by $\theta_v$ is not too large, and that the
energy-deficit effect at the other side of the jet is not too small.
We have also found that the main features in the power-law model are
also applicable to some other angular energy distributions, e.g.
such as the Gaussian form (\ref{Gauss}).

For ease of discussion, we have here assumed that the upper limit of
validity of the assumed angular distribution is $\Theta>2\theta_v$.
This is to avoid that the observer feels the energy deficit beyond
$\Theta$ before the relativistic beaming angle exceeds $\theta_v$. Indeed,
if $\Theta < 2\theta_v$, $\bar\Gamma(\theta_v)$ starts to deviate from
the value predicted by the adiabatic law $\propto t^{-3/8}$ after it
is less than $(\Theta-\theta_v)^{-1}$. In this regime, the upper limits
for $\theta$-integration in the numerators of both (\ref{epsbar}) and
(\ref{epsbar2}) should be replaced by $\Theta$. Thus the maximum
correction factor with respect to the $\Theta > 2\theta_v$ case is a
factor of $\Theta/2\theta_v$. Even for $\Theta=\theta_v$ (i.e., the
line of sight marginally cuts the jet edge), the deviation is at most
a factor of 1/2. We therefore conclude that the $\Theta$ effect may
in most cases not be important. The main reason is that the large
angles contribute a small portion of the total energy in the beam due
to the distribution of the form (\ref{eps}).

In our model, the ``isotropic'' luminosity function will be determined
by the assumed angular distribution, $N(\epsilon)d\epsilon=N(\theta)
d\theta \propto \sin \theta d\theta \propto \theta d\theta$ (the
latter being for small $\theta$). From equations (\ref{eps}),
(\ref{Gauss}) and substituting $\epsilon$ by $L$, we get the
luminosity function predictions in our model, e.g., 
\be
N(L) dL \propto L^{-1-2/k} dL
\label{NLdL1}
\ee
for the power-law model, and
\be
N(L) dL \propto L^{-1} dL
\label{NLdL2}
\ee
for the Gaussian model. To test these luminosity functions, redshift
measurements are needed. Using only the bursts for which optical 
redshifts have been determined so far, e.g. as compiled in F01, PK01,
the above luminosity distributions are not consistent. However, this 
discrepancy could be due to small number statistics ($\sim 20$ in all
or $\sim 10$ on each side of the mid-point). The fact that the small 
sample size or other selection effects related to the afterglow detections 
could lead to a spurious inconsistency is also suggested, for example, 
by the clear deficit of low luminosity (e.g. possibly due to large 
viewing angle) bursts at higher redshifts ($z>1$) in F01's data set.
Alternatively, some other distance indicators have been proposed, such as 
spectral time-lags, e.g. Norris, Marani \& Bonnell 2000, or variability 
measures, e.g.  Fenimore \& Ramirez-Ruiz 2002, Reichart et al. 2001 (and 
interpretations of these indicators in terms of the viewing angle have 
been discussed by, e.g., Salmonson \& Galama 2002 and Norris et al 2002). 
If one accepts such distance indicators and their inferred redshifts at 
face value, the observational GRB luminosity function inferred for a much 
larger bursts sample (e.g. Schaefer, Deng \& Band 2001) is not inconsistent 
with the theoretical distribution (\ref{NLdL1}).  And, using a different 
approach, Schmidt (2001) obtained a flatter observational GRB luminosity 
function, which over a large range of luminosities is compatible with the 
model distribution (\ref{NLdL2}). 

A natural consequence of this model is that the distribution of break 
times $t_b$, and hence the $\theta_v$ distribution, should be related
to the statistical distribution of viewing angles and to the shape of 
the beam distribution. The present data and the preliminary calculations 
are not sufficient to draw firm constraints on parameters of such models.
However, the comparison of such predictions or more detailed versions of 
them against future data in the {\em Swift} era should provide interesting
constraints, as a larger quantity of more accurate redshift measurements 
become available.

\acknowledgements
We are grateful to M.J. Rees for discussions which stimulated this
research, to E. Rossi for sending us their paper, and to
B. Paczy\'nski, J. Granot, E. Waxman, L. J. Gou for valuable comments
or discussions. This work is supported by NASA (NAG5-9192 and
NAG5-9153).

\end{document}